**Understanding how a falling ball chain can be speeded up by impact onto a surface**

J. Pantaleone, Department of Physics, University of Alaska Anchorage, Anchorage Alaska, 99508

When a falling ball chain strikes a surface, a tension is created that pulls the chain downward. This causes a downward acceleration that is larger than free-fall, which has been observed by recent experiments. Here a theoretical description of this surprising phenomenon is developed. The equation of motion for the falling chain is derived, and then solved for a general form of the tension. The size of the tension needed to produce the observed motion is relatively small and is explained here as coming from the rotation of a link just above where the chain collides with the surface. This simple model is used to calculate the size of the tension in terms of physically measurable quantities: the length and width of a link, the maximum bending angle at a junction, the inclination angle of the surface, and the coefficients of friction and restitution between the chain and the surface. The model's predictions agree with the results of current experiments. New experiments are proposed that can test the model.

**1. Introduction.**

Chains are common in both the natural world and also in man-made technology. They exist at the microscopic scale in the form of polymers and at the macroscopic scale on our bicycles. Chains act like smooth, continuous systems on large scales but their discreteness is apparent on small scales. Chains have many similarities with granular media, such as sometimes resembling a fluid and sometimes a solid (see e.g. [1,2]). The properties of chains are of interest for possible industrial and technological applications, as well as for expanding our knowledge of fundamental science.

There are many different types of chains, but one particular type has been the subject of many recent experiments: ball chains. These consist of hollow spheres connected by short rods. Such chains have been used to understand the properties of polymers (see e.g. [3-6]) and they have also been used to highlight some of the counterintuitive dynamical properties of chains. For example, observations of ball chains falling from a container have found that the chain can spontaneously rise up above the container, a phenomenon called a "chain fountain" [7-11]. Also, when a ball chain lying on a table is pulled horizontally, the chain can spontaneously rise up above the table [12,13]. In addition, a ball chain falling onto a surface can accelerate faster than free-fall---the surface literally sucks the chain downward [14,15] (see also [16,17]). These latter phenomena are examples of dynamical instabilities, where an effect created at a small scale grows to manifest itself on much larger scales.

In this article the focus will be on the case of a falling ball chain being pulled towards a surface by its interaction with the surface. Falling ropes and chains have been a topic of discussion since at least as early as 1815 as examples of 1D continuous mass systems [18-22]. However the greater than free-fall acceleration of such systems towards a surface was not reported until 2010 when Hamm and Geminard showed that ball chains exhibited this extra acceleration [14]. This surprising result was reinforced in 2011 by Grewal, Johnson and Ruina [16,17] who used a different type chain, one constructed of stiff, almost horizontal links. This chain exhibited the extra downward acceleration caused by impact with a surface and, unlike for the ball chain, the mechanism for the downward acceleration was readily apparent. When one end of a single, falling, almost horizontal, rod strikes a surface, it causes a rotation of the rod that increases the downward acceleration of the other end of the rod. In a chain, this extra downward acceleration of the lowest rod is spread throughout the chain, making the downward acceleration greater than free-fall.

A series of experiments on this topic was performed recently by Corbin, Hanna, Royston, Singh and Warner [15]. They dropped ball chains onto surfaces inclined at different angles and compared the observed motion to that of an identical chain in free fall. They measured $\Delta y$, the vertical separation of these two chains. These authors found that the chains falling onto an inclined surface fell faster, except at one intermediate angle when the two chains fell at the same rate. Also, there was an important difference between chains falling onto steep surfaces compared to those falling onto relatively flat surfaces. When falling onto steeply inclined surfaces $\Delta y$ increased smoothly throughout the fall, however when falling onto a surface with a shallow inclination a vertical separation only appeared at the later stages of the fall. These experimental results provide important insight into the nature of the force acting on the bottom of the chain and they will be analyzed and explained in this article.

A priori, the case of the ball chain falling onto a highly inclined surface presents a particularly straightforward case for theoretical analysis. This is because then the experiment with the ball chain appears similar to the chain of Grewal, Johnson and Ruina [16,17]. Collision with the surface creates rotation of the links, which then produces the tension in the chain above. Motivated by the new experiments, the present work attempts to understand and explain the experimental results.

This paper is divided into several parts. First is a description of the physical forces acting on the chain as a whole and a derivation of the chain's equation of motion. Then an analytical solution for the vertical motion is derived for the general form of the forces acting on the chain. Next a microscopic model of the chain's interaction with a surface is developed. This model is then used to derive quantitative values for the tension acting on the end of the chain. Using this model, the predicted motion of the chain is compared to the observed motion. Finally, several new experiments are suggested that can test the model.

## 2. The equation of motion for the chain.

Fig. 1 shows the physical situation under consideration. The chain is assumed to initially be vertical, with no horizontal motion, and is falling onto a surface inclined at an angle $\theta$ with respect to the vertical. The forces acting on the falling chain are the weight of the chain above the surface, $mg$, and a force on the lower end of the chain, $(T_x, T_y)$, produced by its interaction with the inclined surface. The weight of the chain can be written as $mg$

= $\lambda yg$, where $\lambda$ is the mass per unit length of the chain and $y$ is the length of the chain above the surface. Observations show that the tension created in the chain acts towards the surface, as indicated by the arrows in the figure.

The literature describing the motion of falling chains can be rather confusing. To clarify the situation, explicit details are given here on how to derive the equation of motion. Each small chunk of the chain with mass $\delta m$ is assumed to obey Newton's second law of motion, $F = \delta m\, a$. At a given instant in time $F = \delta m\, a$ is summed over each piece of the chain above the point of contact with the surface. Because the chain is essentially rigid for displacements along its length, the vertical acceleration of each piece of the chain is the same, so the acceleration factors out of the sum. This leaves a sum over the mass chunks $\delta m$, which yields the total mass, $\lambda y$. In the sum over the forces the internal forces cancel out, by Newton's third law, leaving only the total external forces acting on the chain. Thus the equation of motion in the vertical direction for the chain above the point of contact with the surface is

$$\lambda y \frac{d^2 y}{dt^2} = -\lambda y g - T_y \qquad (1)$$

where $T_y$ is the downward force acting on the end of the chain from its interaction with the surface.

A previously published article [14] derived an equation of motion for a falling chain equivalent to Eq. (1) by starting with $F = dp/dt$ (where $p = m\, dy/dt$ is the momentum) as the more fundamental form of Newton's second law. However $F = dp/dt$ is generally not the correct way to treat variable mass systems, see e.g. Refs. [23-25]. Doing so in the above analysis would add an extra term of $-\lambda(dy/dt)^2$ on the right-hand side of Eq. (1). Including such a term in Eq. (1) is obviously incorrect---then the equation would fail for the case of a chain in free-fall (when $d^2y/dt^2 = -g$, $T_y = 0$, but the term $-\lambda(dy/dt)^2$ would be nonzero and variable). Eq. (1) can also be derived from the general equation for a variable mass system (see e.g. [25]) and is the correct equation of motion for a chain falling onto a surface.

The y-component of the tension force can approximately be parameterized as

$$T_y \approx \alpha \lambda \left[ \frac{dy}{dt} \right]^2 \qquad (2)$$

where $\alpha$ is a small, dimensionless parameter. This form is implicit in some previous theoretical analyses of falling chains [14,20-22] and was justified therein based on macroscopic considerations, such as dimensional analysis. Here a microscopic model is developed in section 4 that explicitly shows how the form in Eq. (2) emerges and calculates $\alpha$ in terms of the physical properties of a link. Also, it should be noted that one obvious problem with Eq. (2) is that it does not account for the fact that the tension at the free end of the chain is always zero, so $T_y$ must vanish when $y$ goes to zero. The microscopic model developed in section 4 will explicitly show why this occurs and why it is not important for the experiments presently under discussion.

A notation different than Eq. (2) was used in refs. [14,15,19-22]. These authors followed the notation of ref. [14], which focused on the vertical component of the force produced by the chain on a horizontal surface beneath it, $F_y$. This force can be related to the tension in Eq. (2) by examining the forces acting on the chain pile that occurs below a chain falling onto a horizontal surface. Applying the general equation for a variable mass system (see e.g. [25]) to this chain pile gives

$$T_y + F_y - \lambda s g = \lambda \left[\frac{dy}{dt}\right]^2 \tag{3}$$

The left hand side of Eq. (3) is the net force upwards on the chain pile; the tension, $T_y$, plus the force from the horizontal surface, $F_y$, minus the weight of the chain in the pile, $\lambda s g$, where $s$ is the length of chain in the pile. The right-hand side of Eq. (3) is the change in vertical momentum flux at the top of the pile. This term occurs when there is a relative velocity between the incoming mass (here the falling chain) and the center of mass of the system (in this case the chain pile). Substituting Eq. (2) into Eq. (3), and solving for $F_y$, there is a part of $F_y$ that is described by $(1-\alpha)\lambda(dy/dt)^2$. Reference [14] called this part the "dynamical force" and parameterized it using the dimensionless constant $\gamma=1-\alpha$. While previous authors used the parameter $\gamma$, here the parameter $\alpha$ is preferred. This is because $\alpha$ describes the tension, which is the "extra" effect that was not included in historical analyses of falling chains. Also, in recent experiments [15], the surface is no longer always horizontal and $F_y$ is not measured. The experimental results in references [14,15] yield $\alpha << 1$, which has important implications for many aspects of the chain's motion.

For example, let's consider the effects of the horizontal force $T_x$ on the horizontal motion of the chain. The chain is not rigid in the horizontal direction so a horizontal force will create a local transverse displacement of the chain. This displacement will propagate along the chain as a wave. As a first approximation, the speed of transverse waves on the chain should be the same as for waves on a smooth string

$$V_{wave} \approx \sqrt{T_y/\lambda} \approx \sqrt{\alpha}\left|\frac{dy}{dt}\right| \tag{4}$$

where Eq. (2) was used for the tension. Because $\alpha$ is a very small parameter the speed of transverse waves is much less than the speed of the chain, $|dy/dt|$. In fluid dynamics this type of flow is called supercritical flow. Eq. (4) implies that the effects of the horizontal force on the end of the chain will not propagate up the falling chain [14]. This agrees with the experiments in Ref. [15] in that they do not report a significant horizontal displacement of the chain, even though it is to be expected that $T_x > T_y$ for a steeply inclined surface.

The expression for the wave velocity used in Eq. (4) applies to small deviations on a thin, flexible string. There are several aspects of the chain that violate those assumptions and so might allow for some transverse motion to travel faster than described by the flexible string approximation: static rolling friction at the junctions, large angle bending at the junctions, sphere-sphere collisions and the discrete size of the links. Rolling friction is typically a very small effect for rigid solids and will be neglected. The other effects are likely to act only over a relatively short region of the chain near the surface, with Eq. (4) applying above that small region. These latter effects will be somewhat taken into account in our model for how the tension force is produced.

## 3. Solving the vertical equation of motion.

Experimental observations of chains falling onto a surface have used video analysis to measure the vertical separation distance between the chain falling onto a surface and an identical chain in free-fall. In this section a simple, analytical equation will be found for the final separation distance, $\Delta y$. This will facilitate comparisons between theory and experiment.

The equation of motion in the vertical direction is obtained by combining Eqs. (1) and (2) to give

$$u\frac{d^2u}{d\tau^2} = -2u - \alpha\left(\frac{du}{d\tau}\right)^2 \tag{5}$$

Here the equation has been written in terms of dimensionless parameters $u = y/y_0$ (where $y_0$ is the initial length of the chain) and $\tau = t/t_f$ (where $t_f = (2y_0/g)^{1/2}$ is the free-fall time starting from rest). This equation can be integrated by using that $d^2u/d\tau^2$ can be written as $\beta d\beta/du$ where $\beta = du/d\tau$. Then the first and last terms in Eq. (5) can be combined to give [15,21]

$$\frac{d}{du}\left[u^{2\alpha}\beta^2\right] = -4u^{2\alpha} \tag{6}$$

Eq. (6) can be integrated from the initial conditions of $u = 1$, $\beta = \beta_0$ to find $\beta$ at an arbitrary position. The resulting equation can then be rearranged to solve for a quadrature for $\tau$. Integrating from the initial condition, $u=1$, to the final condition, $u = 0$, this integral gives the time for the whole chain to fall to the collision point as

$$\tau_f = \int_0^1 du \frac{u^\alpha}{\left\{\beta_0^2 + \left(\frac{4}{1+2\alpha}\right)\left[1 - u^{1+2\alpha}\right]\right\}^{1/2}} \tag{7}$$

This can be integrated [26] to give

$$\tau_f = \frac{\Gamma\left[\frac{1+\alpha}{1+2\alpha}\right]{}_2F_1\left[\frac{1}{2}, 1; \frac{2+3\alpha}{1+2\alpha}; \frac{-4}{(1+2\alpha)\beta_0^2}\right]}{(1+2\alpha)|\beta_0|\Gamma\left[\frac{2+3\alpha}{1+2\alpha}\right]} \tag{8}$$

where $\Gamma$ is the gamma function and ${}_2F_1$ is the hypergeometric function. This expression has a much simpler form for the most common form of the experiment in Ref. [15], the case when the chain is released from the rest with the lower end of the chain just barely touching the surface. Then the initial velocity vanishes so $\beta_0 = 0$ and

$$\tau_f = \frac{1}{2}\sqrt{\frac{\pi}{1+2\alpha}} \frac{\Gamma\left[\dfrac{1+\alpha}{1+2\alpha}\right]}{\Gamma\left[\dfrac{3+4\alpha}{2+4\alpha}\right]} \tag{9}$$

The chain striking the surface is moving faster than the chain in free fall, so $\tau_f$ is less than the time it takes the whole chain in free-fall to reach the collision point. Thus the final separation distance is $u_{\text{free-fall}}(\tau_f)$, the free-fall equation of motion evaluated at time $\tau_f$.

$$\frac{\Delta y}{y_0} = 1 + \beta_0 \tau_f - \tau_f^2 \tag{10}$$

Note that $\beta = (2/y_0 g)^{1/2} \, dy/dt$ is a negative quantity because the chain is moving downward.

The expression for $\tau_f$ and for the separation distance, $\Delta y$, can be further simplified by making use of the fact that $\alpha \ll 1$. Doing a power series expansion gives for the final separation distance, at leading order in $\alpha$,

$$\frac{\Delta y}{y_0} \approx \alpha \left[ -2 - \beta_0\left(\beta_0 + \sqrt{\beta_0^2 + 4}\right) + \frac{1}{4}(\beta_0^2 + 4)\ln[\beta_0^2 + 4] + \frac{1}{2}(\beta_0^2 + 4)\ln\left[\frac{\beta_0^2 + 4 + \beta_0\sqrt{\beta_0^2 + 4}}{-\beta_0 + \sqrt{\beta_0^2 + 4}}\right] \right] \tag{11}$$

This has a particularly simple form when $\beta_0 = 0$.

$$\frac{\Delta y}{y_0} \approx 2\left[2\ln(2) - 1\right]\alpha \tag{12}$$

The experiments have typically found $\Delta y/y$ to be a few percent, so these last two expressions should be reasonable.

As a check, it is worth noting that Eqs. (11) and (12) can be derived directly, and more easily, using a different method. The tension force can be treated as a small perturbation to free-fall motion. Then at leading order in $\alpha$ the separation distance is given by integrations over the free-fall solutions for the position and velocity

$$\frac{\Delta y}{y_0} \approx \alpha \int_0^{\tau_{ff}} d\tau \int_0^{\tau} d\tau' \frac{\left[\beta_0 - 2\tau'\right]^2}{\left[1 + \beta_0 \tau' - \tau'^2\right]} \tag{13}$$

Here $\tau_{ff} = \left[\sqrt{4 + \beta_0^2} + \beta_0\right]/2$ is the time is takes the chain in free-fall to go from the initial to final position, $u = 1$ to $u = 0$. Integration of Eq. (13) gives Eq. (11) and, in the limit $\beta_0 \to 0$, Eq. (12).

The dependence of the final separation distance, $\Delta y$, on the initial chain speed is illustrated in Fig. 2. There the analytical solution given by Eqs. (8) and (10) is used to make a plot of $\Delta y/(\alpha y_0)$ versus the magnitude of the initial velocity, $|\beta_0|$, for three different values of $\alpha$. Fig. 2 shows that $\Delta y/(\alpha y_0)$ is relatively insensitive to $\alpha$ for $\alpha \ll 1$,

as is to be expected for any regular function that vanishes as $\alpha = 0$. Somewhat more surprisingly, Fig. 2 shows that $\Delta y/(\alpha y_0)$ is also relatively insensitive to the initial velocity, $|\beta_0|$, varying by less than 25% as the velocity changes from 0 to infinity for these $\alpha$ values. This behavior agrees with the results of the experiments in Ref. [15]. Fig. 2(left) in Ref. [15] plots $\Delta y$ versus drop height, corresponding to a change in $|\beta_0|$ from 0 to 1.8. The analytical solution predicts only a small increase of $\Delta y/y_0$ over this range, about 16%, in agreement with the data in Ref. [15].

**4. Modeling the tension as a function of inclination angle.**

**A) Identifying the types of collisions with the surface, and the location of the transition region between them.**

When a falling ball chain strikes an inclined surface, what happens depends on the inclination of the surface. For almost horizontal surfaces the falling chain will collect in a pile, with the spheres in the chain colliding with other spheres in the chain. In contrast, for almost vertical surfaces the spheres of the chain will each impact once with the surface and then will smoothly move away from the collision point, with no sphere ever touching another sphere in the chain. We start by estimating the angle of the inclination corresponding to the transition between these two extreme cases.

When the surface is steeply inclined, it is reasonable to assume that the chain is moving smoothly away from the collision point, and not collecting at the collision point, even if the surface is somewhat rough. The rate of mass flowing into the collision point must equal the rate flowing out

$$\lambda_{in} V_{in} = \lambda_{out} V_{out} \qquad (14)$$

where $\lambda$ is the linear mass density and $V$ is the chain speed. The speed of the chain out of the collision region is slower than the speed into the collision region because mechanical energy is lost during the collision with the surface. Thus Eq. (14) implies that the density of the chain, $\lambda$, must increase after the collision. This can occur easily for a ball chain because it is made of hollow spheres connected by short rods (that are flared on the ends inside the spheres). When compressed the rods slide into the spheres, increasing the chain's linear mass density. However this type of density increase cannot continue forever, the compression fundamentally changes when the outgoing spheres start to touch each other. Then the chain flowing out of the collision region will buckle. Using Eq. (14), the condition for when buckling starts to occur in the outgoing flow can be written as

$$\frac{V_{out}}{V_{in}} = \frac{D}{L} \qquad (15)$$

where $D$ is the diameter of the spheres and $L$ is the distance between the centers of the spheres for a stretched chain, see Fig. 3. This relation determines the critical slope for the outgoing flow to buckle.

A different condition determines when buckling occurs in the incoming chain. Buckling there starts when the sphere rebounding from the surface collides with the sphere above it in the falling chain. The critical condition for when this occurs is when the lower sphere has a grazing impact with the upper sphere.

$$\frac{V'_y}{V'_x} = \frac{\sqrt{L^2 - D^2}}{D} \tag{16}$$

Here $V'_x$ and $V'_y$ are the horizontal and vertical components of the velocity of the lower sphere after the collision, in the reference frame where the incoming chain is at rest, see Fig. 3 inset.

In order to determine the critical inclination angles corresponding to the conditions in Eqs. (15) and (16), the velocity of the chain just after the collision with the surface must be known. As a first approximation, we shall calculate the velocity of the sphere just leaving the collision point with the surface assuming that it is free, with no external forces acting on it besides those associated with the interaction with the inclined surface during the collision. This is reasonable because the chain is compressed downstream of the colliding sphere and, after the collision, the bottom-most link is compressed in the upstream direction too. When the chain is compressed the spheres are relatively independent of each other so the force from the surrounding spheres should be small. In this approximation the velocity of the sphere after the collision with the surface can be written as (in the falling chain's rest frame).

$$\begin{aligned} V'_x &= V \sin\theta (1+\varepsilon)[\cos\theta - \mu \sin\theta] \\ V'_y &= V \sin\theta (1+\varepsilon)[\sin\theta + \mu \cos\theta] \end{aligned} \tag{17}$$

Here $V$ is the velocity of the chain above the collision point, $\varepsilon$ is the coefficient of restitution (the ratio of the sphere's speed perpendicular to the surface after the collision to the same component before the collision) and $\mu$ is the coefficient of kinetic friction between the sphere and the surface (see e.g. reference [27]). The angle $\theta$ is with respect to the vertical, as shown in Figs. 1 and 3, while the angle with respect to the horizontal (used for the abscissa in Fig. 4) is 90-$\theta$. Eq. (17) assumes that the sphere slides during the interaction with the surface, so it is not a complete description of all possibilities because it does not include the change in the friction force that occurs when sliding stops and rolling begins. However Eq. (17) describes the primary effects of a friction force and so is appropriate given the crude approximation of assuming the sphere is "free".

Substituting Eq. (17) into Eq. (16) gives an equation for the critical angle in the incoming flow. In the ratio of velocity components, the incoming chain speed, $V$, and the coefficient of restitution, $\varepsilon$, cancel out so that the critical angle only depends on the coefficient of friction, $\mu$. To evaluate the expression for the critical angle we specialize to the case of the experiments in reference [15] and use that $(D/L)^2 \approx 0.5$, which holds for a wide variety of ball chains [28], and also assume that $\mu \ll 1$ since the chain and inclination surfaces used in these experiments are smooth. Then the critical angle can be written as

$$\theta_{C,i} \approx \frac{\pi}{4} - \mu - \mu^2 \tag{18}$$

where $\theta$ is in radians and the expression is accurate to order $\mu^3$. The subscript on $\theta$ in Eq. (18) denotes that this is the critical angle for buckling in the incoming flow.

The critical angle for the outgoing flow to buckle is obtained by shifting the velocities in Eq. (17) to the lab reference frame, where the inclined surface is at rest, and then substituting them into Eq. (15). We again specialize to the measurements of Ref. [15], taking $(D/L)^2 \approx 0.5$ and $\mu \ll 1$. Also, the videos of chain collisions with the surface show that the flow after the collision point is mostly along the surface so we additionally assume that $\varepsilon \ll 1$. Then the critical angle can be written as

$$\theta_{C,o} \approx \frac{\pi}{4} - \mu + \frac{1}{2}\varepsilon^2 - \mu\varepsilon - \frac{1}{2}\mu^2 \qquad (19)$$

where $\theta$ is in radians and the expression is accurate to order small parameters cubed. The subscripts on $\theta$ denote that this is the critical angle for buckling of the outgoing chain.

Comparing the expressions for the two critical angles, Eqs. (18) and (19), we see that they are the same at leading order and also at first order. Thus, for the condition in Ref. [15], it is reasonable to expect that buckling will occur in the outward flow at approximately the same angle as it occurs for the incoming flow. This conclusion is also reasonable on more general grounds because once buckling occurs in either the incoming or outgoing flow, it alters the direction of the sphere leaving the surface and would tend to increases the likelihood of buckling occurring in the other flow.

For the experimental observation of chains falling as a function of inclination angle [15], values for the parameters $\varepsilon$ and $\mu$ are not given. However values for these parameters can be estimated from what information is given. From the close-up videos of scattering at angle of $\theta = 90° - 67.5° = 22.5°$, the change in velocity from the collision with the surface can be measured and fit using the parameterization in Eq. (17) to give $\mu = 0.20 \pm 0.07$ and $\varepsilon = 0.07 \pm 0.02$. This corresponds to a critical angle of $\theta_C \approx 34° \pm 4°$, with buckling occurring at larger angles (shallow inclinations). This value is consistent with statements in Ref. [15] that the transition from coiling (a type of buckling) to "smooth" flow for the incoming chain is stated as occurring for $\theta$ in the range 45° to 37.5 degrees, and with videos in Ref. [15] showing that rebounding spheres collide with the incoming spheres at 37.5 degrees, but do not come near the sphere above at the next smallest angle where a video is given, 22.5°. The critical angle is indicated in Fig. 4 at the angle with respect to the horizontal of $90° - \theta_C \approx 56° \pm 4°$.

**B) How collision with a surface causes tension**

Consider the motion of the two spheres above the collision point as shown in Fig. 3. The bottom sphere is colliding with the surface and it experiences a force from the surface. This force will cause the bottom sphere to move upwards, and horizontally, and to rotate. The upward force on the bottom sphere will not give rise to a similar upward force on the top sphere because the rod connecting the two spheres can move into the spheres. However the connecting rods can still transmit a torque to the upper sphere. This torque will cause the upper sphere to rotate about a pivot at, or near, the top of the sphere. This rotation produces a downward force on the chain above the upper sphere from the centripetal motion of the center of mass of the sphere. This tension is approximately

$$T_y \approx M\langle\omega^2\rangle\frac{D}{2} \qquad (20)$$

where $M$ is the mass of the sphere, $\omega$ is the angular velocity and $D$ is the diameter of the sphere. The brackets are there to denote that the instantaneous force needs to be averaged over time to agree with how the tension is defined in Eq. (2)

This expression for the tension, Eq. (20), contains several simplifying assumptions. It assumes that the maximum rotation angle is small, so that the centripetal force is approximately the same as its downward component. This is reasonable since the maximum rotation angle at a junction for the ball chain in Ref. [15] is $\phi_{max} = 47°/2 \approx 23°$. Also, it assumes that the mass of the chain above the pivot is large compared to the mass of a single sphere, so that the pivot point will undergo a negligible displacement in the vertical direction in the chain's center of mass rest frame. In addition, it assumes that the pivot point is fixed in the horizontal direction as well. This should be relatively reasonable when the mass of the chain above the pivot is large, given the general idea expressed by Eq. (4), that transverse motion is not transmitted upwards along the chain. Finally, Eq. (20) neglects the fact that as the sphere rotates the geometry of the junction is such that the rod moves into the sphere, thus the location of the pivot will change with the rotation angle. This could lead to significant (factor of 2) modifications of Eq. (20) at large rotation angles. Accounting for all of these details would mask the basic physics behind the creation of the tension and will not be pursued here. Instead Eq. (20) will be used to capture the essential physics of how the tension is created, while bearing in mind its quantitative limitations.

While the sketch in Fig. 3 is for a steeply inclined surface, the same general mechanism applies for a surface that is only slightly inclined. In this case the horizontal motion of the chain is generated indirectly by the buckling of the chain near the collision region. In the buckling region, sphere-sphere collisions creates a direct, upwards force between the spheres that compresses the chain, causing the buckling. This buckling involves only a very small region of the chain at the bottom because the upper part of the chain is moving supercritically, see Eq. (4). Thus Eq. (20) will apply at the top of the buckling region.

The assumption that the pivot is fixed in the falling chain's rest frame will break down as the mass of the chain above it goes to zero. In this limit there is no fixed pivot that constrains the motion of the sphere, so the force on this point will vanish. This explains the observation made in the discussion below Eq. (2), that $T_y$ must vanish as $y$ goes to 0. The model above indicates that this effect is only relevant when the mass of the chain above the pivot point is comparable to the mass of a single sphere. For the experiments under consideration this section of the chain is very small compared to the total length of the chain. Thus a decrease in the force over this very small region will have little effect on the $\Delta y$ calculated in the previous section. This effect will be neglected for the current application.

The time average in Eq. (20) can be reduced to the time interval over which the falling chain is periodic. Since the uncompressed chain is periodic with a length $L$, and moving with a speed $V$, this averaging is over the timescale

$$t_{collision} = \frac{L}{V} \qquad (21)$$

The evaluation of Eq. (20) for the two extreme cases, steep and shallow surface inclination, will be treated separately.

**C) Tension for slightly inclined surfaces.**

When the chain falls onto a slightly inclined surface, there is a small region above the collision point where buckling occurs. Buckling is a common phenomenon in several systems. Buckling can take the form of coiling in solid and liquid ropes [29-32], and coiling behavior has been reported for the ball chain system also [15]. For solid and liquid ropes the forces of elasticity or viscosity, respectively, limit and smooth the motion of the rope. However for a ball chain these forces do not apply and the only effect limiting the amplitude of the buckling is the maximum bending angle at a junction. Thus it is to be expected that in the buckling region the transverse displacements are at the maximum possible bending angle at a sphere-rod junction, $\phi_{max}$. At this bending angle a junction "locks", so that the chain is effectively rigid in the bending plane. For a rigid system the wave speed is much faster than the wave speed given in Eq. (4), thus the flow is only supercritical above the buckling region.

To make a crude evaluation of Eq. (20), we shall assume that the change from compression to tension occurs at the top of the buckling region. There the rotation of a sphere is assumed to start from rest and proceeds with constant angular acceleration to the angular value found in the buckling region, $\phi_{max}$, over the time $t_{collision}$. Then

$$T_y \approx \lambda V^2 \frac{4}{3} \frac{D/2}{L} \phi_{max}^2 \qquad (22)$$

where it has been used that $\lambda = M/L$, and the numerical factor of 4/3 follows from the assumption of constant angular acceleration. The tension given in Eq. (22) has the general form assumed in Eq. (2). It is independent of the inclination angle, $\theta$, and also the coefficients of friction and restitution, $\varepsilon$ and $\mu$. This estimate applies for shallow surface inclinations, which correspond to the left-hand side of Fig. 4.

To compare the prediction of Eq. (22) with the experimental results in Ref. [15], the parameter values of $D = 2.4$ mm, $L = 3.36$ mm and $\phi_{max} = 23°$ were used. This gives $\alpha = 0.08$ which, when substituted into Eq. (12) for a total chain length of 0.497 m, corresponds to the horizontal line on the left-hand side of Fig. 4. The estimate in Eq. (22) does not account for the characteristic delay time in the onset of the tension force reported in [15]. Adding a delay time/length would lower the predicted $\Delta y$. Given the crudeness of the estimate there is good agreement with the observations.

The assumption in Eq. (22) that the sphere accelerates smoothly through a rotation angle of $\phi_{max}$ inherently assumes that at the top of the buckling region the chain has a relatively constant curvature. Such a region of coordinated motion will take some time/distance to develop. This distance will be of order the length of chain it takes to form a loop. For the experiments in Ref. [15] this is (1 bead/47°)*360° ≈ 8 beads (where 47° is the maximum bending angle at a bead for the chain used), which explains the characteristic delay length of about 11 beads, independent of inclination angle, found for chains falling onto a slightly inclined surface. Also, it is to be expected that this coherent region is likely to be relatively easy to disrupt even after it has developed. This is because the sphere-sphere scattering that occurs in the buckling region is a classic chaotic process exhibiting sensitivity to initial conditions. Disruption of the coherent region would tend to decrease the rotation angle and so decrease $\Delta y$. This intrinsic irregularity is probably the source of the large fluctuations in the observed chain separation on the left-hand side of Fig. 4.

**D) Tension for steeply inclined surfaces.**

To evaluate Eq. (20) for a steeply inclined surface, it will be assumed that the outgoing flow is approximately parallel to the surface ($\varepsilon \ll 1$). Then the collision with the surface causes the flow direction to change by an angle $\theta$ from its original direction. This means that the angular velocity around the pivot of the two-sphere system shown in Fig. 3 is very quickly changed from zero to approximately $\omega \approx \theta/t_{collision}$. After a subsequent internal collision between the two spheres and the rod connecting them "locks" this system into a single rigid system, the upper sphere will have approximately this same angular velocity around the pivot. Then, using the expression for $t_{collision}$ given in Eq. (21), the tension in Eq. (20) can be written as

$$T_y \approx \lambda V^2 \frac{D/2}{L} \theta^2 f(\theta) \tag{23}$$

Here it has been used that $\lambda = M/L$ and $f(\theta)$ is defined to be the fraction of the collision time during which the upper sphere is rotating.

The function $f(\theta)$ is difficult to estimate. It depends on when the rod connecting the two spheres "locks" at the internal junctions, which starts the rotation of the upper sphere. When this occurs depends on not just the relative velocity between the two spheres but also on the relative initial angular orientation between the two spheres, the angular velocities of the two spheres about their centers, and the distance between the two spheres. Modeling all these elements is beyond the scope of the present work. As a crude estimate, Eq. (23) is evaluated using $f \approx 0.5$ to give the solid squares on the right-hand side of Fig. 4.

In general, the model in Eq. (23) is successful in that it has the form of Eq. (2), has the correct limiting behavior (vanishing at $\theta = 0$), and describes the approximate size of the tension. However the model's predictions for $\Delta y$ are significantly lower than the actual observed values at the smallest $\theta$ (far right side of Fig. 4). This is most likely due to some of the more severe approximations made by Eq. (20).

**E) Tension near the critical angle.**

The tension in the chain is observed to vanish around the critical angle, see Fig. 4. To understand this let's consider approaching the critical angle from the steeply inclined side so that Eq. (23) should be roughly applicable. In this formalism a vanishing tension can occur if $f(\theta) \approx 0$. There are two effects that might make this happen.

First, consider that when the chain is bent into a curve, the angle between neighboring spheres increases as the distance between the spheres decreases. Thus when the spheres are the closest together, just barely touching, the rod-sphere joints have their largest range of motion before locking occurs. Now the critical angle is where the spheres of the chain are at their closest, and if the internal junctions don't lock, no rotation of the upper sphere occurs, and the tension vanishes. Whether this is the case or not depends on many factors that are difficult to calculate without a detailed simulation of the many different parts.

Another factor to be considered is that, at the critical angle, direct contact between the spheres starts to occur, in both the incoming and outgoing flow. Buckling in the outgoing flow will reduce the range of motion of the bottom sphere in Fig. 3, reducing the amount of rotation that occurs in the upper sphere, reducing the tension. Also, buckling in the incoming flow will give rise to a direct, upward force on the upper sphere in Fig. 3, cutting-off the induced tension from that sphere.

In general, the two effects mentioned probably act together. The failure of the internal junctions to lock eliminates the tension from the first part of the bottom spheres motion, while the buckling eliminates the tension from the latter part of the motion during $t_{\text{collision}}$.

## 5. Summary and discussion

A chain falling onto a surface has a tension pulling it downward that is approximately described by $T_y \approx \alpha \lambda V^2$. Only a very small tension force ($\alpha \ll 1$) is needed to explain the results of experiments. An exact, analytical solution of the motion was obtained, Eqs. (8) and (10), valid for arbitrary initial velocity. The solution takes a particularly simple form for small $\alpha$ and zero initial velocity, Eq. (12).

A model was developed to calculate the tension parameter, $\alpha$, in terms of physical parameters: the dimensionless ratio of the diameter to the spacing between spheres ($D/L$), the inclination angle of the surface ($\theta$), the maximum bending angle at a junction ($\phi_{\max}$), and the coefficients of friction ($\mu$) and restitution ($\varepsilon$). In this model the tension arises from transverse motion in a small region of the chain just above the surface. This transverse motion is constrained by the mass of the chain above it, causing a rotation around the end of the chain. It is this rotation which creates the tension. The model was applied to the observations in Ref. [15], and the results generally agree with the observations, see Fig. 4.

In the pioneering work of Hamm and Geminard [14], they qualitatively explained the production of the tension as following from the chain's finite radius of curvature. Since a chain's finite radius of curvature follows from the finite bending angle at a junction, which Eq. (22) relies on, there is a qualitative connection between the two explanations.

Similarly, the tension has also been previously described as resulting from a dissipative shock [20-22]. There is a connection between the present work and those works also since the speed of the collision point is larger than the speed of waves on the chain, see Eq. (4). However, while describing the collision with the surface as a shock does constrain the general form of the tension force on long length scales, it does little to reveal the physical mechanism responsible for the tension. The current analysis goes beyond those works in that it gives a transparent physical mechanism for the production of the tension, and calculates the tension force in terms of the chain's physical parameters.

The model for the creation of the tension force that is proposed here in Eq. (20) is very simplistic. A more quantitatively accurate description would necessarily involve many more degrees of freedom. However the intent here is to sacrifice quantitative accuracy for a clear understanding of the basic principles behind the observed effects. The model provides a framework for discussing the current experimental results and, it is hoped, a guide for future experimental and theoretical work.

The model proposed here makes many testable predictions:

*Buckling starts to occur at the critical angle*. The Experiments in Ref. [15] did not resolve when buckling was occurring, only when coiling (one type of buckling) was occurring. The model predicts that for small coefficients of restitution, $\varepsilon \ll 1$, buckling should start in both the incoming and outgoing flows at the critical angle.

*The critical angle should depend on the coefficient of friction*. Eqs. (18) and (19) tell us that as the coefficient of friction between the chain and the surface is increased, the critical angle should move to steeper inclinations.

*The size of $\Delta y$ should increase as the coefficient of restitution, $\varepsilon$, is increased, for a steeply inclined surface*. The coefficient of restitution depends on the materials the chain and the base are made of, so in principle this parameter can be varied. While Eq. (23) assumes $\varepsilon \approx 0$, the logic behind it predicts that $\Delta y$ depends on the square of the change in angle the chain makes from its initial direction. Increasing $\varepsilon$ will increase this change in angle, and so will also increase $\Delta y$. For example, for $\mu \ll 1$ and $\varepsilon \approx 1$ the change in angle would be $2\theta$, so $\Delta y$ would be 4 times larger.

*The size of $\Delta y$ should decrease as $\phi_{max}$ decreases for a shallowly inclined surface*. While the ratio $D/L$ is the same for many different chains from the same manufacturer [28], the maximum bending angle, $\phi_{max}$, varies considerably between different sized ball chains from the same manufacturer [33]. Eq. (22) predicts that $\Delta y$ depends on $\phi_{max}$ squared, so this should be an easily observable effect.

In addition to the experiments in Ref. [15], the tension in a chain falling onto a flat surface has been measured in a couple of other experiments. The original experiment of Hamm and Geminard [14] measured the force that a ball chain exerted onto a flat surface and found $\alpha \approx 0.17$ with no characteristic delay time reported. This value for $\alpha$ is larger than the maximum value of 0.11 found in Ref. [15].   The difference in these results might be due to a difference in experimental uncertainties between the two different experimental methods.  However the model in Eq. (22) suggests another possible explanation, the two experiments may have used chains with different values of $\phi_{max}$.  The value for this parameter was not given in Ref. [14] so a direct comparison of the two experimental results is not possible.

Information on a collision-induced tension can also be obtained from chain fountain experiments.  These experiments typically observe the motion of a long chain moving from a container onto a horizontal surface.  The steady-state speed the chain moves at depends on the force acting on the chain as it strikes the surface.  This force can be directly determined by measuring the chain speed and the distance from the surface to the highest part of the chain.  The one experiment so far to make these measurements [10] found $\alpha \approx 0$.  Chain fountain experiments differ fundamentally from the previous two measurements of $\alpha$ in several important ways.  In particular, the chain does not strike the horizontal surface in the same location all of the time, but instead the impact location varies erratically over a region orders of magnitude larger than the diameter of the chain.  This large horizontal variation may prevent the creation of a coherent structure near the top of the buckling region, effectively preventing the creation of a tension force.  Stated differently, the erratic horizontal motion may expand the characteristic length where $\alpha = 0$ (observed in Ref. [15]) for when the tension force develops.  More experiments are needed to resolve this issue.

The model developed here to describe the creation of the tension has similarities to the model used to explain the creation of the "push-off" force responsible for the chain fountain [8-10].  For both phenomena it is the rotation of a link of the chain that produces a force.  However there is a fundamental difference between the two mechanisms.  For the chain fountain the link is being stretched while for impact with a surface the link is being compressed.  Compression means it is harder for the joints in the link to lock, making the tension sensitive to the parameter $\phi_{max}$.  In contrast, for the chain fountain the link is stretched and the created force is less sensitive to this parameter [10,33].

**Data accessibility statement.**  Data used to support the content of this work can be found in the publications given in the reference list.

**Competing interests statement.**  I have no competing interests.

**Authors' contribution statement.**  I am the sole author.

**Funding.**  I have received no funding specifically for this work.  However I thank the University of Alaska Anchorage for providing me time to do research.

**Ethics statement.**  This is a theoretical work and did not involve experimentation on humans, animals, rocks or fossils.

**Acknowledgements.**  I thank James Hanna for bringing his work in Ref. 15 to my attention.

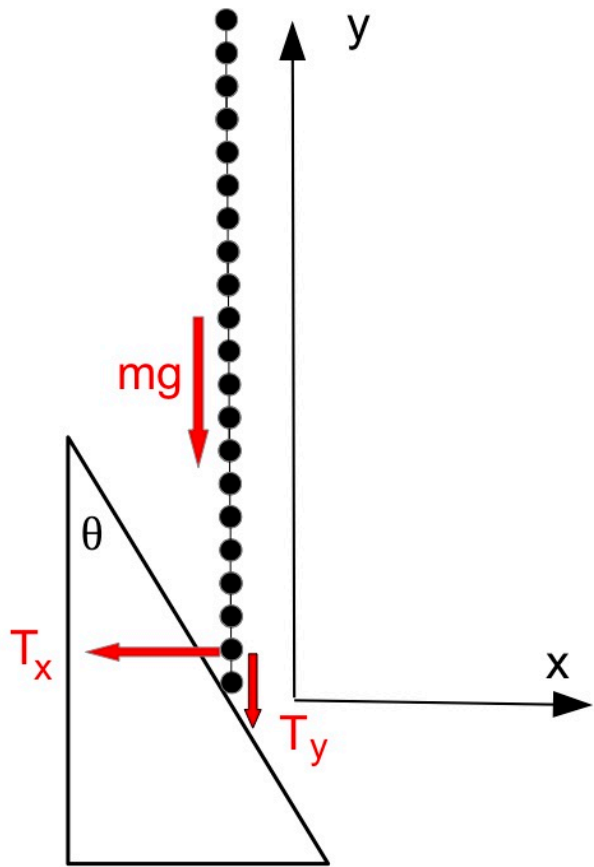

Fig. 1. Sketch of the forces acting on the falling chain. The force components $T_x$ and $T_y$ describe the tension created near the bottom of the chain, just above where the chain collides with the surface.

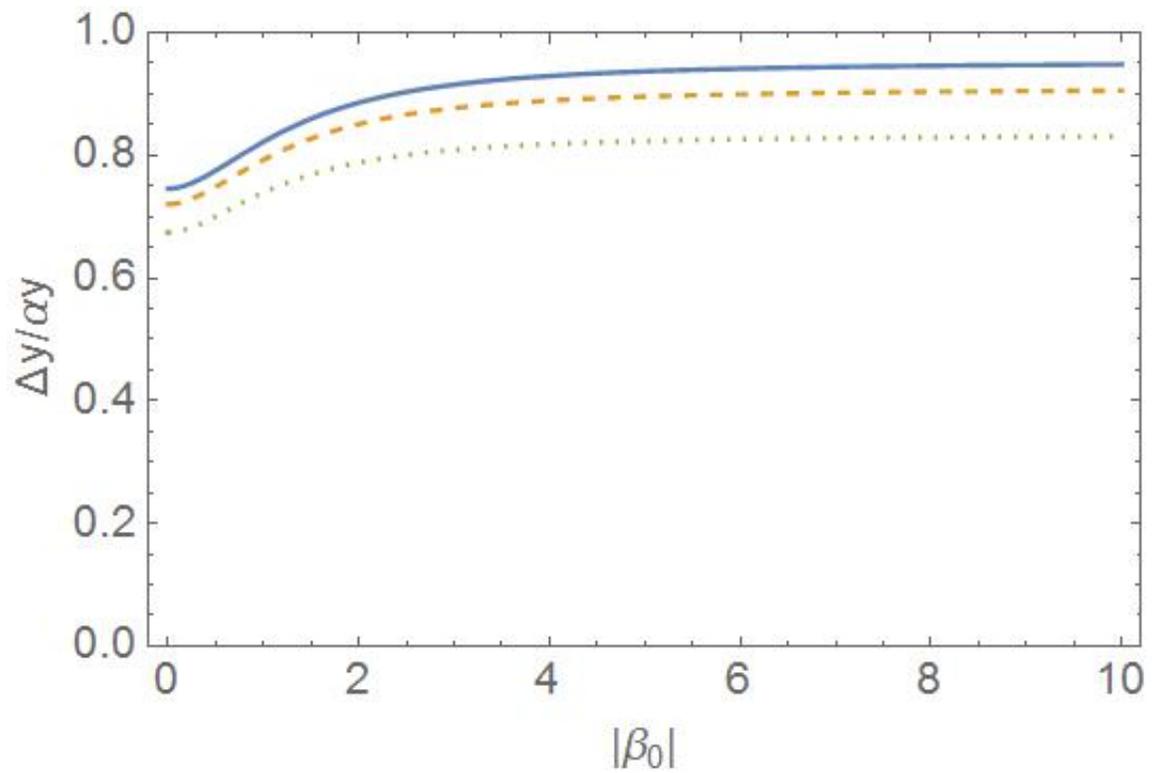

Fig. 2. Plot of the relative separation distance divided by dimensionless tension parameter, $\Delta y/(\alpha y_0)$, versus the absolute value of the initial velocity of the chain striking the surface, $|\beta_0|$, for $\alpha = 0.05$ (solid), 0.10 (dashed) and 0.20 (dotted).

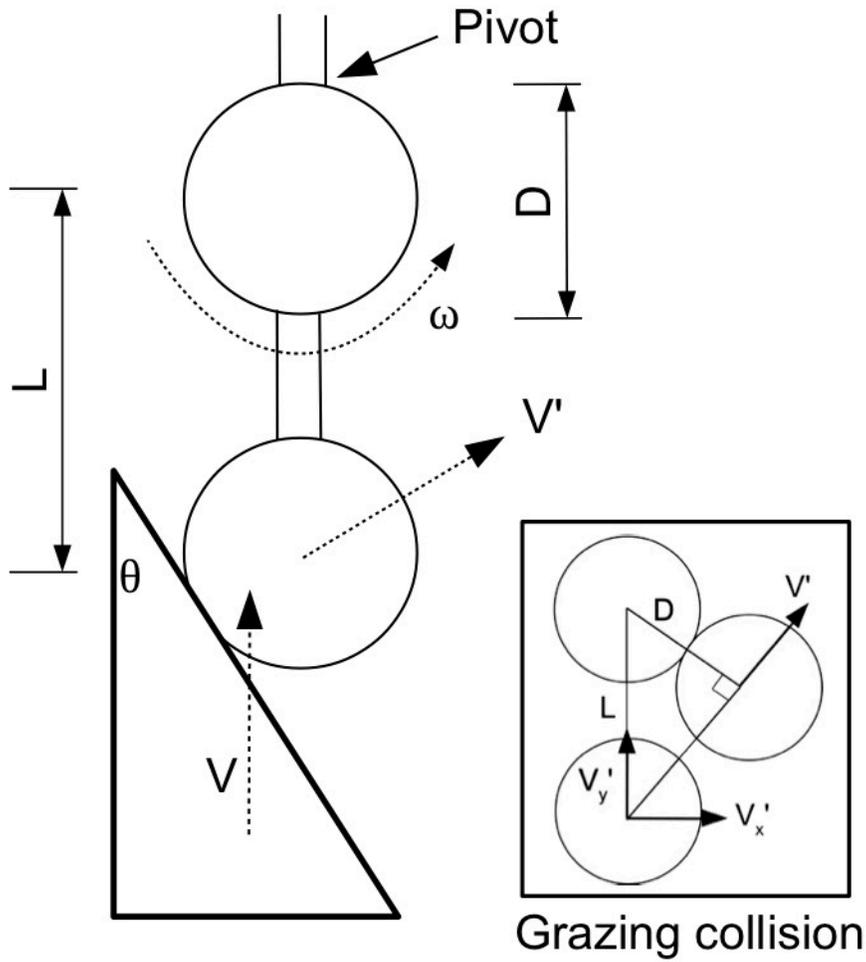

Fig 3. Sketch showing the motion of the bottom pieces of the chain as viewed in the chain rest frame. In this reference frame the surface is moving upward with speed V, collides with the lower sphere and transfers some energy to it, which causes a rotation in the upper sphere. This rotation produces a tension in the chain above it. The inset shows the geometry for the onset of buckling in the incoming flow, as described by Eq. (16).

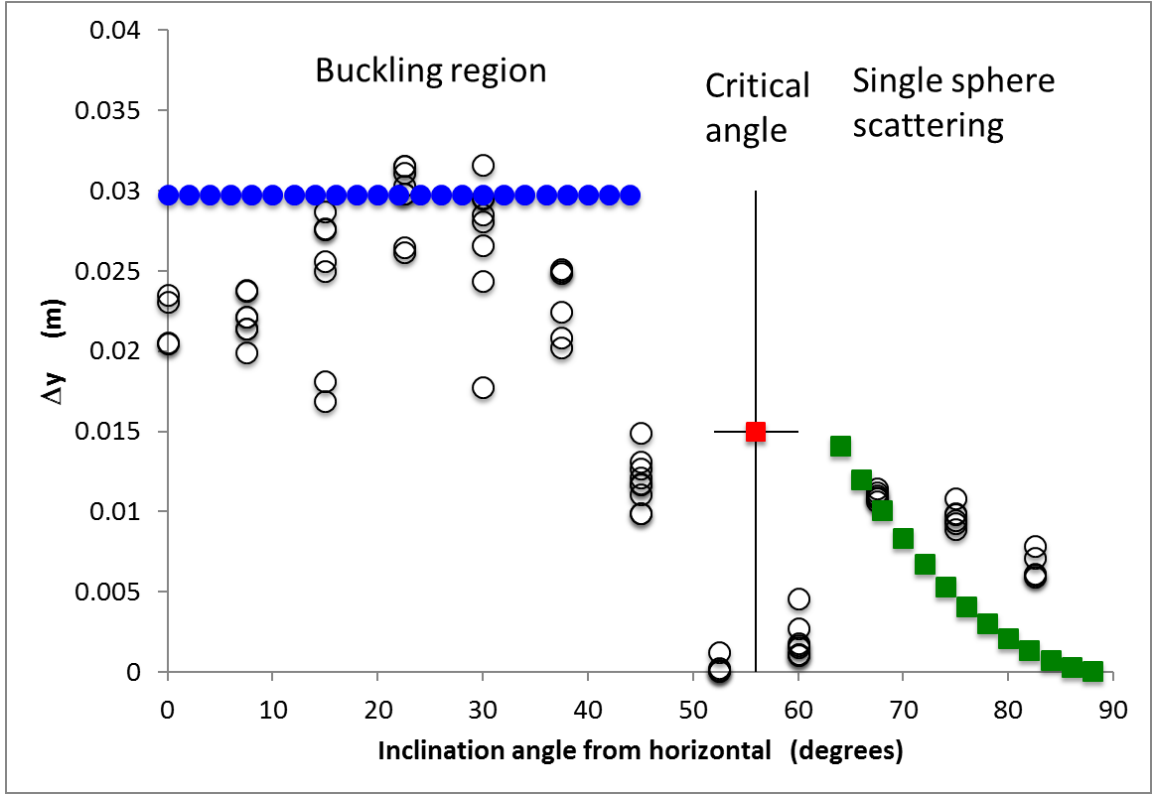

Fig. 4. Comparison of theory and experiment for the final chain separation, $\Delta y$, versus surface angle with respect to the horizontal $(90-\theta)$. Open circles correspond to observations reported in Ref. [15]. Solid circles on the left are the predictions from Eq. (22) and the solid squares on the right are predictions using Eq. (23). The vertical line (around the red square) is the predicted critical angle, calculated using Eq. (18) and a coefficient of friction obtained from videos in the single sphere scattering region.